\documentstyle[12pt]{article}
\newcommand{\vf}{\varphi}
\newcommand{\be}{\begin{equation}}
\newcommand{\ee}{\end{equation}}
\newcommand{\ba}{\begin{eqnarray}}
\newcommand{\ea}{\end{eqnarray}}
\newcommand{\no}{\nonumber\\}
\newcommand{\bi}{\bibitem}
\begin{document}
\thispagestyle{empty}
\begin{flushright}
IFA-FT-392-1993 \\
December 1993
\end{flushright}
\bigskip\bigskip\begin{center}
{\bf \Huge{THE GEODESIC MOTION IN}}
\end{center}
\begin{center}
{\bf \Huge{TAUB-NUT SPINNING SPACE}}
\end{center}
\vskip 1.0truecm
\centerline{{\bf\Large
{Mihai Visinescu\footnote{E-mail address:~~ MVISIN@ROIFA.BITNET}}}}
\vskip5mm
\centerline{Department of Theoretical Physics}
\centerline{Institute of Atomic Physics, P.O.Box MG-6, Magurele,}
\centerline{Bucharest, Romania}
\vskip 2cm
\bigskip \nopagebreak \begin{abstract}
\noindent
The geodesic motion of pseudo-classical spinning particles in the
Euclidean Taub-NUT space is analysed. The generalized Killing
equations for spinning space are investigated and the constants of
motion are derived in terms of the solutions of these equations. A
simple exact solution, corresponding to trajectories lying on a cone,
is given.
\end{abstract}

\newpage\setcounter{page}1
\section{Introduction}
Spinning particles can be described by pseudo-classical mechanics
models involving anti-commuting c-numbers for the spin-degrees
of freedom. The configuration space of spinning particles (spinning space)
is an extension of a (pseudo-) Euclidean manifold, parametrized by local
co-ordinates {$\{$}$x^\mu${$\}$}, to a graded manifold parametrized by local
co-ordinates {$\{$}$x^\mu, \psi^\mu${$\}$}, with the first set of variables
being Grassmann-even ( commuting ) and the second set Grassmann-odd
(anti-commuting) [1-9].

In spite of the fact that the anti-commuting Grassmann variables do
not admit a direct classical interpretation, the Lagrangians of these
models turn out to be suitable for the path-integral description of
the quantum dynamics. The pseudo-classical equations aquire physical
meaning when averaged over inside the functional integrals [1,10].
In the semi-classical regime, neglecting higher-order quantum
correlations, it should be allowed to replace some combinations of
Grassmann spin-variables by real numbers. Using these ideas the motion
of spinning particles in external fields have been studied in Refs.
[1, 11-13].

On the other hand, generalizations of Riemannian geometry based on
anti-commuting variables have been proved to be of mathematical interest.
Therefore the study of the motion of the spinning particles in curved
space-time is well motivated.

In the present paper we investigate the motion of pseudo-classical
spinning point particles in the Euclidean Taub-NUT space. The Kaluza-Klein
monopole of Gross and Perry [14] and of Sorkin [15] was obtained by
embedding the Taub-NUT gravitational instanton into five-dimensional
Kaluza-Klein theory. Remarkably the same object has re-emerged in the
study of monopole scattering. In the long distance limit, neglecting
radiation, the relative motion of the monopoles is described by the
geodesics of this space [16,17]. Slow Bogomolny-Prasad-Sommerfield
monopoles move along geodesics in the Euclidean Taub-NUT space. The
dynamics of well-separated monopoles is completely soluble, but not
trivial [16-22]. The problem of geodesic motion in this metric has
therefore its own interest, independently of monopole scattering.

For all of these reasons, the extension of the Euclidean Taub-NUT
space with additional fermionic dimensions, parametrized by vectorial
Grassmann co-ordinates {$\{$}$\psi^\mu${$\}$} follows naturally. At last the
spinning Taub-NUT space would be a relevant manifold to investigate
the properties of the Killing-Yano tensors [9].

The plan of this paper is as follows. In Sec. 2 we summarize the
relevant equations for the motions of spinning points in curved
spaces. For a systematic treatment of these subjects, the reader is
referred to the original literature, and, expecially, to Refs. [6-9].
In Sec. 3 we analyze the motion of pseudo-classical spinning
particles  in the Euclidean Taub-NUT space. We examine the generalized
Killing equations for this spinning space and describe the derivation of
constants of motion in terms of the solutions of these equations. The
contribution of the spin to the angular momentum is contained in the
Killing scalars. In Sect. 4 we solve the equations given in the previous
Section for the special case of motion on a cone. This case represents an
extension of the scalar particle motions in the usual Taub-NUT space in
which the orbits are conic sections [18-21]. An explicit exact solution
is given. In spite of its simplicity, this solution is far from trivial.
Our comments and concluding remarks are presented in Sec. 5.
\section{Motion in spinning space}
An action for the geodesics of spinning space is:
\be
 S=\int_{a}^{b}d\tau \left(\,{1\over 2}\,g_{\mu \nu}(x)\,\dot{x}^\mu
\,\dot{x}^\nu\, +\, {i\over 2}\, g_{\mu \nu}(x)\,\psi^\mu \,{D\psi^\nu\over
D\tau} \right).
\ee
Here and in the following the overdot denotes an ordinary proper-time
derivative $d/d\tau$, whilst the covariant derivative of $\psi^\mu$
is defined by
\be
{D\psi^\mu\over D\tau}=\dot{\psi}^\mu+\dot{x}^\lambda \,\Gamma^\mu
_{\lambda\nu}\,\psi^\nu .
\ee

The trajectories, which make the action stationary under arbitrary
variations $\delta x^\mu $ and $\delta \psi^\mu$ vanishing at the end
points, are given by:
\ba
{D^2 x^\mu\over D\tau^2}&=&\ddot x^{\,\mu} + \Gamma^\mu_{\lambda\nu}
\,\dot{x}^\lambda\,\dot{x}^\nu={1\over 2i}\,\psi^\kappa\,\psi^\lambda
\,R^{~~\mu~}_{\kappa\lambda~\nu}\,\dot{x}^\nu\\
{D\psi^\mu\over D\tau}&=&0 .
\ea

The anti-symmetric tensor
\be
S^{\mu\nu}=-i\,\psi^\mu\psi^\nu
\ee
can formally be regarded as the spin-polarization tensor of the particle
[1-8]. The equations of motion can be expressed in terms of this tensor
and in particular eq.(4) asserts that the spin is covariantly constant
\be
{DS^{\mu\nu}\over D\tau}=0 .
\ee

The concept of Killing vector can be generalized to the case of spinning
manifolds. For this purpose it is necessary to consider variations of
$x^\mu$ and $\psi^\mu$ which leave the action $S$ invariant modulo
boundary terms. Let us assume the following forms of these variations:
\ba
\delta x^\mu&=&{\cal R}^\mu (x, \dot x, \psi)=R^{(1)\mu}(x, \psi)+
\sum_{n=1}^{\infty} {1\over n!}\dot x^{\nu_1}\cdots \dot x^{\nu_n}
R^{(n+1)~\mu}_{\nu_1 \cdots \nu_n}(x, \psi) \no
\delta\psi^\mu&=&{\cal S}^\mu (x, \dot x, \psi)=S^{(0)\mu}(x, \psi)+
\sum_{n=1}^{\infty} {1\over n!}\dot x^{\nu_1}\cdots \dot x^{\nu_n}
S^{(n)~\mu}_{\nu_1 \cdots \nu_n}(x, \psi)
\ea
and the Lagrangian transforms into a total derivative
\be
\delta S =\int_{a}^{b} d\tau {d\over d\tau}\left(\delta x^\mu\, p_\mu
-{i\over 2}\, \delta \psi^\mu \,g_{\mu\nu}\, \psi^\nu- {\cal J}(x, \dot x,
\psi)\right)
\ee
where $p_\mu$ is the canonical momentum conjugate to $x^\mu$
\be
p_\mu=g_{\mu\nu}\dot{x}^\nu+{i\over 2}\Gamma_{\mu\nu;\,\lambda}
\,\psi^\lambda\psi^\nu=\Pi_\mu + {i\over 2}\Gamma_{\mu\nu;\,\lambda}
\,\psi^\lambda\psi^\nu
\ee
$\Pi_\mu$ being the covariant momentum.

{}From Noether's theorem, if the equations of motion are satisfied,
the quantity ${\cal J}$ is a constant of motion.

If we expand ${\cal J}$ in a power series in the covariant momentum
\be
{\cal J}(x, \dot x, \psi) ={\cal J}^{(0)}(x, \psi) +
\sum_{n=1}^{\infty}{1\over n!}\Pi^{\mu_1}\cdots \Pi^{\mu_n}
{\cal J}^{(n)}_{\mu_1 \cdots \mu_n}(x, \psi)
\ee
then ${\cal J}$ is a constant of motion if its components satisfy a
generalization of the Killing equation [6, 7] :
\be
{\cal J}^{(n)}_{(\mu_1 \cdots \mu_{n};\,\mu_{n+1})}+ {\partial
{\cal J}^{(n)}_{(\mu_1\cdots \mu_n}\over\partial\psi^\sigma}
\Gamma^{~~~~~~\sigma}_{\mu_{n+1})\lambda}\ \psi^\lambda=
{i\over 2}\,\psi^\sigma \,\psi^\lambda\, R_{\sigma\lambda\nu(\mu_{n+1}}
{\cal J}^{(n+1)\nu}_{\mu_1\cdots\mu_n)} .
\ee

In general the symmetries of a spinning-particle model can be divided
into two classes.  First, there are conserved quantities which exist
in any theory and these are called {\it generic} constants of motion.
The second kind of conserved quantities, called {\it non-generic},
depend on the explicit form of the metric $g_{\mu\nu}(x)$.

In Refs. [6, 7] it was shown that for a spinning particle model
defined by the action (1) there are four generic symmetries:
\begin{enumerate}
\item{Proper-time translations and the corresponding constant of
motion is the Hamiltonian
\be
H={1\over 2}\, g^{\mu\nu}\,\Pi_\mu\,\Pi_\nu
\ee}
\item{Supersymmetry generated by the supercharge
\be
Q=\Pi_\mu\,\psi^\mu
\ee}
\item{Chiral symmetry generated by the chiral charge
\be
\Gamma_{*}={1\over 4!}\sqrt{-g}\epsilon_{\mu\nu\lambda\sigma}
\,\psi^\mu\,\psi^\nu\,\psi^\lambda\,\psi^\sigma
\ee}
\item{Dual supersymmetry, generated by the dual supercharge
\be
Q^{*}={1\over 3!}\,\sqrt{-g}\epsilon_{\mu\nu\lambda\sigma}
\,\Pi^\mu\,\psi^\nu\,\psi^\lambda\,\psi^\sigma  .
\ee}
\end{enumerate}
In the next Section we shall apply these results and we shall discuss
specific solutions of the generalized Killing equations to the case of
a spinning particle moving in a Euclidean Taub-NUT space.
\section{Euclidean Taub-NUT spinning space }
The Kaluza-Klein monopole [14,15] was obtained by embedding the Taub-NUT
gravitational instanton into five-dimensional theory, adding the time
coordinate in a trivial way. Its line element is expressed as:
\ba
ds^2_5&=&-dt^2+ds^2_4\no
&=&-dt^2+V^{-1}(r)[dr^2+r^2d\theta^2+r^2\sin^2\theta\,d\vf^2]\no
& &+V(r)[dx^5+\vec{A}(\vec{r})\,d\vec{r}\,]^2
\ea
where $\vec{r}$ denotes a three-vector $\vec{r} =(r, \theta, \vf)$ and
the gauge field $\vec{A}$ is that of a monopole
\ba
A_r=A_\theta&=&0,~~~A_{\vf}=4m(1-\cos\theta)\no
\vec{B}&=&rot\vec{A}={4m\vec{r}\over r^3} .
\ea

The function V(r) is
\be
V(r)=\left(1+{4m\over r}\right)^{-1}
\ee
and the so called NUT singularity is absent if $x^5$ is periodic with
period $16\pi m$ [23].

It is convenient to make the co-ordinate transformation
\be
4m(\chi+\vf)=-x^5
\ee
with $0\leq \chi < 4\pi$, which converts the four-dimensional line
element $ds_4$ into
\be
ds^2_4=V^{-1}(r)[dr^2+r^2d\theta^2+r^2\sin^2\theta\, d\vf^2]
+16m^2 V(r)[d\chi+\cos\theta\, d\vf]^2 .
\ee

Spaces with a metric of the form given above have an isometry group
$ SU(2)\times U(1)$. The four Killing vectors are
\be
D^{(\alpha)}=R^{(\alpha)\mu}\,\partial_\mu,~~~~\alpha=1,\cdots ,4
\ee
where
\ba
D^{(1)}&=&{\partial\over\partial\chi}\no
D^{(2)}&=&-\sin\vf\,{\partial\over \partial\theta}-\cos\vf\,\cot\theta
\,{\partial\over\partial\vf}+{\cos\vf\over\sin\theta}\,{\partial\over
\partial\chi}\no
D^{(3)}&=&\cos\vf\,{\partial\over \partial\theta}-\sin\vf\,\cot\theta
\,{\partial\over\partial\vf}+{\sin\vf\over\sin\theta}\,{\partial\over
\partial\chi}\no
D^{(4)}&=&{\partial\over\partial\vf}.
\ea

$D^{(1)}$ which generates the $U(1)$ of $\chi$ translations, commutes
with the other Killing vectors. In turn the remaining three vectors
obey an $SU(2)$ algebra with
\be
[D^{(2)}, D^{(3)}]=-D^{(4)}~~~,{\it etc... }
\ee

This can be contrasted with the Schwarzschild space-time where the
isometry group at spacelike infinity is $SO(3)\times U(1)$. This
illustrates the essential topological character of the magnetic
monopole mass [24].

In the purely bosonic case these invariances would correspond to
conservation of the so called "relative electric charge" and the
angular momentum [18-22] :
\be
q=16m^2\,V(r)\,(\dot\chi+\cos\theta\,\dot\vf)
\ee
\be
\vec{j}=\vec{r}\times\vec{p}\,+\,q\,{\vec{r}\over r} .
\ee

The first generalized Killing eq.(11) has the form
\be
B^{(\alpha)}_{~~,\,\mu}+{\partial B^{(\alpha)}\over\partial
\psi^\sigma}\,\Gamma^\sigma_{\mu\lambda} \,\psi^\lambda=
{i\over 2}\,\psi^\rho\psi^\sigma\, R_{\rho\sigma\lambda\mu}
\,R^{(\alpha)\lambda}
\ee
which shows that with each Killing vector $R^{(\alpha)}_\mu$
there is an associated Killing scalar $B^{(\alpha)}$.
Therefore if we limit ourselves to variations (7) to
terminate after the terms linear in $\dot x^\mu$, we obtain the
constants of motion
\be
J^{(\alpha)}=B^{(\alpha)} + \dot x^\mu R^{(\alpha)}_\mu
\ee
which asserts that the Killing scalars contribute to the "relative
electric charge" and the total angular momentum.

Inserting the expressions for the connections and the Riemann curvature
components corresponding to the Taub-NUT space in eq.(26) after a
long and tedious calculation we obtain for the Killing scalars:

\ba
B^{(1)}=&-&{32m^3\cos\theta\over (4m+r)^2}\,S^{r\vf}-
{32m^3\over(4m+r)^2}\,S^{r\chi}+{8m^2 r\sin\theta\over 4m+r}\,S^{\theta\vf}
\no
{}~\no
B^{(2)}=& &2m\sin\vf \,S^{r\theta}-{16m^2r+2mr^2\over (4m+r)^2}
\sin\theta\cos\theta\cos\vf \,S^{r\vf}\no
&-&{4m(8m^2 +8mr+r^2)\over (4m+r)^2}\sin\theta\cos\vf \,S^{r\chi}\no
&-&\left[ {8m^2 r\over 4m+r}\sin^2\theta\cos\vf+{4mr(2m+r)\over
4m+r}\cos^2\theta\cos\vf\right] \,S^{\theta\vf}\no
&-&{4mr(2m+r)\over 4m+r}\cos\theta\cos\vf \,S^{\theta\chi}+
{4mr(2m+r)\over 4m+r}\sin\theta\sin\vf \,S^{\vf\chi}\no
{}~\no
B^{(3)}=&-&2m\cos\vf \,S^{r\theta}-{16m^2r+2mr^2\over (4m+r)^2}
\sin\theta\cos\theta\sin\vf \,S^{r\vf}\no
&-&{4m(8m^2 +8mr+r^2)\over (4m+r)^2}\sin\theta\sin\vf \,S^{r\chi}\no
&-&\left[ {8m^2 r\over 4m+r}\sin^2\theta\sin\vf+{4mr(2m+r)\over
4m+r}\cos^2\theta\sin\vf\right] \,S^{\theta\vf}\no
&-&{4mr(2m+r)\over 4m+r}\cos\theta\sin\vf \,S^{\theta\chi}-
{4mr(2m+r)\over 4m+r}\sin\theta\cos\vf \,S^{\vf\chi}\no
{}~\no
B^{(4)}=& &\left[(2m+r)\sin^2\theta+{32m^3\cos^2\theta\over (4m+r)^2}
\right] \,S^{r\vf}+{32m^3\cos\theta\over (4m+r)^2} \,S^{r\chi}\no
&+&{8mr^2+r^3\over 4m+r}\sin\theta\cos\theta \,S^{\theta\vf}-
{8m^2r\over 4m+r}\sin\theta \,S^{\theta\chi} .
\ea

Taking into account the contribution of the Killing scalars to eq.
(27) one finds for the constants of the motion ${\cal J}^{(\alpha)}$ :
\ba
{\cal J}^{(1)}=&B^{(1)}&+q\no
{}~\no
{\cal J}^{(2)}=&B^{(2)}&-(4m+r)r\sin\vf\cdot\dot{\theta}-r(4m+r)
\sin\theta\,\cos\theta\,\cos\vf\cdot\dot{\vf}\no &+&q\,\sin\theta
\,\cos\vf \no
{}~\no
{\cal J}^{(3)}=&B^{(3)}&+(4m+r)r\cos\vf\cdot\dot{\theta}-r(4m+r)
\sin\theta\,\cos\theta\,\sin\vf\cdot\dot{\vf}\no &+&q\,\sin\theta
\,\sin\vf \no
{}~\no
{\cal J}^{(4)}=&B^{(4)}&+r(4m+r)\sin^2\theta\cdot\dot{\vf}+
q\,\cos\theta.
\ea

We remark that the "relative electric charge " $q$ is no longer
conserved contrasting with the purely bosonic case. On the other
hand the conserved total angular momentum is the sum of the orbital
angular momentum, the Poincar\'e  contribution and the spin angular
momentum:
\be
\vec{{\cal J}}=\vec{B}+{\vec j}
\ee
with $\vec{\cal J}=({\cal J}^{(2)}, {\cal J}^{(3)}, {\cal J}^{(4)}) $
and $\vec{B}=(B^{(2)}, B^{(3)}, B^{(4)})$.

{}From eqs. (26) we can derive an interesting linear combination of
${\cal J}^{(\alpha)},\\  \alpha =1,\dots ,4 $ :
\ba
{\cal J}^{(1)}-{\vec{{\cal J}}\cdot\vec{r}\over r}=&-&r\sin^2\theta
\,\cos\theta \,S^{r\vf}+4m\sin^2\theta \,S^{r\chi}\no
&-&r^2\,\sin\theta\,\cos^2\theta
\,S^{\theta\vf}+4mr\,\sin\theta\,\cos\theta \,S^{\theta\chi} .
\ea

In the standard Taub-NUT space, eq.(31) reduces to
\be
{\vec{j}\cdot\vec{r}\over r}=|\vec{j}| \cos\theta = q
\ee
which fixes the relative motion to lie on a cone whose vertex is at
the origin and whose axis is $\vec{j}$. Eq. (31) express the fact
that the total angular momentum in the radial direction receives
contributions from the spin angular momentum, the orbital angular
momentum being absent in that direction. Moreover the motion is more
complicated than in the bosonic case since in general the angle
$\theta$ is not constant in time.

In addition to these constants of motion there are four universal
conserved charges described in the previous Section. Using the notation
from this Section they are:

1.The energy
\ba
E=&{1\over 2}&V^{(-1)}(r)\left[\dot{\vec{r}}^{\,2} +\left( {q\over
4m}\right)^2\right]\no
&=& {1\over 2} {4m+r\over r}\,\dot r^2
+{1\over 2}(4m+r)\,r\,\dot\theta^2 +{1\over 2}(4m+r)\,r\,\sin^2\theta\,
\dot\vf^2\no
& &+ 8m^2\,{r\over 4m+r}\,(\cos\theta\,\dot\vf +\dot\chi)^2
\ea

2.The supercharge
\ba
Q=& &{4m+r\over r}\dot r\,\psi^r+(4m+r)\,r\dot\theta\,\psi^\theta\no
&+&\left[ (4m+r)\,r\sin^2\theta\,\dot\vf+q\,\cos\theta\right]\,\psi^\vf
+q\,\psi^\chi
\ea

3.The chiral charge
\be\Gamma_* =4m(4m+r)r\,\sin\theta\,\psi^r\,\psi^\theta\,\psi^\vf
\,\psi^\chi
\ee

4.The dual supercharge
\ba
Q^*&=&4m(4m+r)r\,\sin\theta(\dot r\,\psi^\theta\,\psi^\vf\,\psi^\chi-
\dot\theta\,\psi^r\,\psi^\vf\,\psi^\chi+\dot\vf\,\psi^r\,\psi^\vf
\,\psi^\chi  \no
& &-\dot\chi\,\psi^r\,\psi^\theta\,\psi^\vf) .
\ea

Finally, having in mind that $\psi^\mu$ is covariantly constant, the
rate of change of spins is:
\ba
\dot\psi^r&=&{2m\over r(4m+r)}\,\dot r\,\psi^r+{r^2+2mr\over 4m+r}
\,\dot\theta\,\psi^\theta\no
& &+\left({r^2+2mr\over 4m+r}\,\sin^2\theta
+{32m^3r\cos^2\theta\over (4m+r)^3}\right)\dot\vf\,\psi^\vf\no
& &+{32m^3r\cos\theta\over(4m+r)^3}(\dot\vf\,\psi^\chi+\dot\chi\,\psi^\vf)
+{32m^3r\over(4m+r)^3}\dot\chi\,\psi^\chi\no
{}~\no
\dot\psi^\theta&=&-{r+2m\over r(4m+r)}(\dot r\,\psi^\theta+
\dot\theta\,\psi^r)+{8mr+r^2\over (4m+r)^2}\sin\theta\,\cos\theta\,\dot\vf
\,\psi^\vf\no
& &-{8m^2\sin\theta\over (4m+r)^2}(\dot\vf\psi^\chi+\dot\chi\psi^\vf)
\no
{}~\no
\dot\psi^\vf&=&-{r+2m\over r(4m+r)}(\dot r\,\psi^\vf+\dot\vf\,\psi^r)
-{8m^2+8mr+r^2\over (4m+r)^2}\,{\cos\theta\over\sin\theta}\,
(\dot\theta\,\psi^\vf+\dot\vf\,\psi^\theta)\no
& &+ {8m^2\over (4m+r)^2}\,{1\over \sin\theta}\,(\dot\theta\,\psi^\chi+
\dot\chi\,\psi^\theta)\no
{}~\no
\dot\psi^\chi&=&{\cos\theta\over (4m+r)}(\dot r\,\psi^\vf+\dot\vf\,\psi^r)
-{2m\over r(4m+r)}(\dot r\,\psi^\chi+\dot\chi\,\psi^r)\no
& &+\left({8m^2+8mr+r^2\over (4m+r)^2}{\cos^2\theta\over\sin\theta}
+{1\over 2}\sin\theta\right)(\dot\theta\,\psi^\vf+\dot\vf\,\psi^\theta)
\no
& &-{8m^2\over (4m+r)^2}\,{\cos\theta\over\sin\theta}\,(\dot\theta\,
\psi^\chi+\dot\chi\,\psi^\theta).
\ea

As a rule these complicated equations could be integrated to obtain
the full solution of the equations of motion for the usual co-ordinates
$\{$$ x^\mu$$\}$ and Grassmann co-ordinates $\{$$ \psi^\mu$$\}$.

For example, it is posible to use an iterative procedure starting with
the motion of a spinless point particle in the Taub-NUT space. Moreover
in this case there is a conserved vector analogous to the Runge-Lenz
vector of the Coulomb problem whose existence is rather surprising in
view of the complexity of the equations of motion. This conserved vector
is [18]:
\be
\vec{{\cal K}}=\vec{p}\times\vec{j}\, +\, \left({q^2\over 4m}-4mE\right)
{\vec{r}\over r}.
\ee

The trajectories lie hence simultaneously on the cone $\vec{j}\cdot
\vec{r}/r=q$ and also in the plane perpendicular to
\be
\vec{n}=q\vec{{\cal K}}+\left( 4mE-{q^2\over 4m}\right)\vec{j}.
\ee
They are thus conic sections.

Starting with a solution of this type, from eqs.(37) it
is possible to find the Grassmann variables $\{$$ \psi^\mu$$\}$ in a
first approximation. The next step is to use eq.(3) to get corrections
to $\{$$ x^\mu$$\}$ due the spin variables and so on. The essential
point is that the iterative procedure brings to an end after a
{\it finite} number of steps taking into account that the co-ordinate
$\{$$ \psi^\mu$$\}$ are anti-commuting variables.

Unfortunately, the equations of motion are quite intricate and the
general solution is by no means illuminating. Instead of the general
solution, in the next Section we shall discuss a special solution
which is very simple and far from trivial.
\section{Special solution}
In this Section we solve the equations given above limiting ourselves
to the motion on a cone. This characteristic of the motion of the scalar
particles in Taub-NUT spaces can be found for spinning particles only in
special cases.

For this purpose let us choose the $z$ axis along $\vec{J}$ so that the
motion of the of the particle may be conveniently described in terms of
polar co-ordinates
\be
\vec{r}=r\vec{e}\,(\theta,\vf)
\ee
with
\be
\vec{e}=(\sin\theta\cos\vf, \sin\theta\sin\vf, \cos\theta).
\ee

Using the iterative procedure described above, we shall start with the
motion of a scalar particle in Taub-NUT space taking $\dot\theta=0$.
Eqs.(37) can be written in terms of the anti-symmetric tensor
$S^{\mu\nu}$ as follows:
\ba
\dot{S}^{r\theta}=&-&{\dot r\over 4m+r}\,S^{r\theta}+{8m+r\over
2(4m+r)^3}\,q\,\sin\theta \,S^{r\vf}-{8m^2\over r(4m+r)^3}\,{q\sin\theta
\over\cos\theta} \,S^{r\chi}\no
&-&{r\sin^2\theta+2m\over(4m+r)^2}\,{q\over\cos\theta}\,S^{\theta\vf}
-{2mq\over (4m+r)^2} \,S^{\theta\chi}\no
{}~\no
\dot{S}^{r\vf}=&-&{q\over2r(4m+r)\sin\theta} \,S^{r\theta}-{\dot r
\over 4m+r}\,S^{r\vf} -{2mq\over (4m+r)^2} \,S^{\vf\chi}\no
{}~\no
\dot{S}^{r\chi}=& &{q\over 2r(4m+r)\sin\theta\cos\theta}\,S^{r\theta}
+{\dot {r}\cos\theta\over 4m+r} \,S^{r\vf}+{(r\sin^2\theta+2m)\,q\over
(4m+r)^2\cos\theta} \,S^{\vf\chi}\no
{}~\no
\dot{S}^{\theta\vf}=& &{(2m+r)\,q\over r^2 (4m+r)^2\cos\theta} \,S^{r\theta}
-2{2m+r\over r(4m+r)}\,\dot {r}\,S^{\theta\vf}+{8m^2\sin\theta \,q\over r
(4m+r)^3\cos\theta} \,S^{\vf\chi}\no
{}~\no
\dot{S}^{\theta\chi}=& &{q\over 8m(4m+r)^2} \,S^{r\theta}+{\dot {r}
\over 4m+r}\cos\theta \,S^{\theta\vf}-{\dot r\over r} \,S^{\theta\chi}
\no
&+&{8m+r\over 2(4m+r)^3}\,q\sin\theta \,S^{\vf\chi}\no
{}~\no
\dot{S}^{\vf\chi}=& &{q\over 8m(4m+r)^2} \,S^{r\vf}-{(2m+r)\,q\over
r^2(4m+r)^2\cos\theta}\,S^{r\chi}\no
&-&{q\over 2r(4m+r)\sin\theta\cos\theta} \,S^{\theta\vf}-{q\over
2r(4m+r)\sin\theta} \,S^{\theta\chi}-{\dot r \over r} \,S^{\vf\chi}.
\ea

Since we are looking for solutions with $\dot\theta =0$ we have
\be
{d\over dt}\left[ {\cal J}^{(1)} -{\vec{{\cal J}}\cdot\vec{r}
\over r}\right] =0
\ee
and from eqs. (31) and (42) we conclude that
\be
S^{r\theta}+ {2r(2m+r)\over 4m+r}\sin\theta \,S^{\vf\chi}=0.
\ee

This relation implies that the special solutions investigated in
this Section are situated in the sector with
\be
\Gamma_* =0.
\ee

Expressing $S^{r\theta}$ through $S^{\vf\chi}$ we can form  the
following combinations which are equivalent with eqs.(42) :
\ba
{d\over dt}\left[ (4m+r) S^{r\vf}\right]&=&{r\over 4m+r}\,q S^{\vf\chi}
\no
{}~\no
{d\over dt}\left[ \cos\theta S^{r\vf}+ S^{r\chi}\right]&=&0
\no
{}~\no
{d\over dt}\left[ r(4m+r) S^{\theta\vf}\right]&=&-2{\sin\theta\over
\cos\theta}\,{r\over 4m+r}\,q \,S^{\vf\chi}
\no
{}~\no
{d\over dt}\left[ r\cos\theta \,S^{\theta\vf}+ r\,S^{\theta\chi}\right]&=&
-{\sin\theta\over 4m}\,{r\over 4m+r}\,q \,S^{\vf\chi}.
\ea

Thus the equations of motion for $ S^{\mu\nu}$ are written in a more
tractable form and the solution follows without difficulties [25].
However the general solution is still quite involved and in what
follows we prefer to present explicitely a particular solution. For
this purpose we shall satisfy eq.(44) in a trivial way:
\be
S^{r\theta}\,=\,S^{\vf\chi}\,=\,0.
\ee

In spite of this drastic simplification, eqs.(46) have a nontrivial
solution:
\ba
S^{r\vf}&=&{{\cal C}^{r\vf}\over 4m+r}\no
S^{r\chi}&=&{\cal C}^{r\chi}\,-\,{\cos\theta\over 4m+r}\,
{\cal C}^{r\vf}\no
S^{\theta\vf}&=&{{\cal C}^{\theta\vf}\over r(4m+r)}\no
S^{\theta\chi}&=&{{\cal C}^{\theta\chi}\over r}\,-\,\cos\theta\,
{{\cal C}^{\theta\vf}\over r(4m+r)}
\ea
where ${\cal C}^{\mu\nu}$ are Grassmann constants. These constants
are not all independent, having two relations between them
\be
{\cal C}^{r\chi} \, =\, {\sin\theta\over 4m}\, {\cal C}^{\theta\vf}
\ee
\be
\sin\theta\cos\theta\,{\cal C}^{r\vf}\, - \, 4m\cos\theta\,{\cal C}^
{\theta\chi}\, - \, 2\sin^2\theta \, {\cal C}^{\theta\vf}\, = \, 0.
\ee

In the case of this particular solution
\be
{\cal J}^{(1)}\,=\,q
\ee
and therefore the "relative electric charge" is conserved as in the
case of the scalar particle in the usual Taub-NUT space. However the
total angular momentum is modified by the spin contribution
\be
{\cal J}^{(1)}\,-\,{\vec{{\cal J}}\cdot\vec{r}\over r}\,=\,
{\cal J}^{(1)}\,-\,{\cal J}\cos\theta\,=\, -\,\sin\theta\,{\cal C}^{
\theta\vf}.
\ee
Here ${\cal J}$ is the magnitude of the total angular momentum and eq.
(52) fixes the angle $\theta$ in terms of the constants $q, {\cal J}$
and ${\cal C}^{\theta\vf}$.

Also the equations for $\vf$ and $\chi$ are modified:
\ba
\dot{\vf}&=&{q\over (4m+r)\,r\cos\theta}\,-\,{1\over (4m+r)^2}\,
{\cal C}^{r\vf}  \,+\,{\sin\theta\over (4m+r)^2\cos\theta}
\,{\cal C}^{\theta\vf}
\no
\dot{\chi}&=&{8m+r\over 16m^2(4m+r)}\,q\,+\,{\cos\theta\over (4m+r)^2}
\,{\cal C}^{r\vf}\,-\,{\sin\theta\over (4m+r)^2}\,{\cal C}^{\theta\vf}.
\ea
Finally,  $\dot r$ can be derived from the energy, eq.(33).

Since for this particular solution the motion of the spinning particle
is restricted to a cone, it is natural to ask for a conserved vector
similar to the Runge-Lenz vector (38). Unfortunately, for the spinning
particles this form turns out to be inadequate.

Motivated by the studies of Peres [26] and Holas and March [27], we
shall construct a vector $\vec{{\cal K}}$ which is constant in time,
as an appropriate generalization of the Runge-Lenz vector (38). Using
the parametrization (40), (41) the velocity vector can be expressed as
\be
\dot{\vec{r}}\,=\,\dot{r}\vec{e}\,+\,r\dot\vf\dot{\vec{e}}\, '
\ee
where
\be
\dot{\vec{e}}\,'\,=\,{d\vec e\over d\vf}\,=\,(-\sin\theta\sin\vf,
\sin\theta\cos\vf, 0 ).
\ee

The generalized Runge-Lenz vector can be written in a local rotating
basis $(\vec{e}(\vf), \vec{e}\,', \vec {j})$ as
\be
\vec{{\cal K}}\,=\, X_1\,\left(\vec{e}\,-\,\cos\theta\,{\vec{j}\over j}
\right)\,+\, X_2\,\vec{e}\,'.
\ee

$\vec{{\cal K}}$ will remain constant in the laboratory frame if it will
rotate in the opposite direction with respect to its local basis
\ba
X_1&=&X_0\,\cos(\vf\,-\,\vf_0)\no
X_2&=&-\,X_0\,\sin(\vf\,-\,\vf_0)
\ea
where $X_0, \vf_0$ are constants which can be choosen as one wishes.

However it is necessary to note that the property of being constant in
time is not sufficient for $\vec{\cal K}$ to be an integral of motion.
It must be a one-valued function of the state of the particle [22, 27].
Therefore, even restricting to the solution described in this Section,
only in some particular cases the orbits are closed paths and the vector
$\vec{\cal K}$ (56) may serve as an integral of motion.
\section{Concluding remarks}
In the last time the pseudo-classical limit of the Dirac theory of a
spin-1/2 particle in curved space-time is described by the supersymmetric
extension of the ordinary relativistic point particle. The spinning space
represents the extension of the ordinary space-time with anti-symmetric
Grassmann variables to describe the spin degrees of freedom.

Our main concern has been the motion of pseudo-classical spinning particles
in Euclidean Taub-NUT space. This space has been considered extensively
in the literature in connection with the study of monopole scattering and
Kaluza-Klein monopole. The geodesic motion in the ordinary Euclidean
Taub-NUT space is integrable and has a remarkably close analogy with motion
under a Coulomb force. The existence of a conserved vector analogous to the
Runge-Lenz vector of the Coulomb problem is rather surprising in view of the
velocity-dependent forces.

In our analysis of the spinning Taub-NUT space we have restricted to the
contribution of the spin contained in the Killing scalars $B^{\alpha}
(x,\,\psi)$ defined by eq.(26). In spite of the complexity of the equations,
we are able to present a special solution which is very simple, but not at
all trivial. Other particular solutions will be presented elsewhere [25].

Extensions of these results to the Killing-Yano tensors [9,28] are possible
and necessary. In general it is desirable to have a deeper understanding of
the role of the Runge-Lenz vector (38) for the motion of spinning particles.
The existence of this vector can be related to a Killing-Yano 2form [19].
On the other hand some properties of the classical Coulomb (Kepler) motion
may be rediscovered in a metric admitting a Killing-Yano tensor of rank four
[28].

These generalizations are under investigations.
\subsubsection*{Acknowledgements}
The author is greatly thankful to N.Dragon and V.I.Ogievetsky for
very useful discussions on the problems of motion of spinning particles
in curved spaces. He wants to thank also D.Baleanu and R.Nicolaescu for
many valuable discussions and collaboration during the preliminary
stages of this work.
\end{document}